\documentclass[11pt]{article}

\usepackage{times}
\usepackage{amssymb,amsmath,amsthm}
\usepackage{mathrsfs}
\usepackage{epsfig}
\usepackage{color}

\def\B{\mathscr B}

\def\C{\mathbb C}
\def\D{\mathscr D}

\def\H{\mathcal H}

\def\N{\mathbb N}

\def\R{\mathbb R}
\def\S{\mathscr S}

\def\v{\varphi}

\def\0{\boldsymbol 0}

\def\12{{\textstyle\frac12}}
\def\<{\left\langle}
\def\>{\right\rangle}
\def\({\left(}
\def\){\right)}
\def\[{\left[}
\def\]{\right]}
\def\dom{\mathcal D}
\def\lone{\mathsf{L}^{\:\!\!1}}

\def\ltwo{\mathsf{L}^{\:\!\!2}}

\def\e{\mathop{\mathrm{e}}\nolimits}
\def\d{\mathrm{d}}

\def\slim{\mathop{\hbox{\rm s-}\lim}\nolimits}

\newtheorem{Theorem}{Theorem}[section]
\newtheorem{Remark}[Theorem]{Remark}
\newtheorem{Lemma}[Theorem]{Lemma}
\newtheorem{Assumption}{Assumption}

\newtheorem{Proposition}[Theorem]{Proposition}\newtheorem{Definition}[Theorem]{Definition}

\begin{document}


\title{Time delay for an abstract quantum scattering process}

\author{S. Richard\footnote{On leave from Universit\'e de Lyon; Universit\'e Lyon 1; CNRS, UMR5208, Institut Camille Jordan, 43 blvd du 11 novembre 1918, F-69622 Villeurbanne-Cedex, France. The author is supported by the Japan Society for the Promotion of Science (JSPS) and by  "Grants-in-Aid for scientific Research".}}

\date{\small}
\maketitle \vspace{-10mm}

\begin{quote}
Graduate School of Pure and Applied Sciences,
University of Tsukuba,\\
1-1-1 Tennodai, Tsukuba,
Ibaraki 305-8571, Japan,\\
E-mail: {\tt richard@math.univ-lyon1.fr}
\end{quote}

\vspace{3mm}

\begin{center}
\emph{Dedicated to Prof. Hiroshi Isozaki on the occasion of this sixtieth anniversary}
\end{center}

\vspace{3mm}


\begin{abstract}
In this short review paper, we discuss the concept of time delay for an abstract quantum scattering system. Its definition in terms of sojourn times is explained as well as its identity with the so-called Eisenbud-Wigner time delay. Necessary and natural conditions for such a construction are introduced and thoroughly  discussed. Assumptions and statements are precisely formulated but proofs are contained in two companion papers written in collaboration with R. Tiedra de Aldecoa.
\end{abstract}


\section{Introduction}\label{Intro}
\setcounter{equation}{0}

Heuristically, the notion of time delay in scattering theory is quite easy to understand.
Given a reference scattering process, this concept should indicate a measure of the advance or of the delay that a system acquires during a slightly different scattering process.
In other words, the time delay should be a measure of an excess or a defect of time that a certain evolution process gets compared to an {\it a priori} process.
The paradigm example of such two related systems consists in a classical particle evolving either freely in an Euclidean space or in the same Euclidean space but under the influence of a compactly supported potential.

Once this general notion is accepted, one might wonder how it can effectively be measured~? For the paradigm example, the traditional setup consists in a series of manipulations: One first considers a family of boxes $B(r)$ centered at the origin and of edges equal to $r>0$. One then measures the time $T_r^0$ spent by the free particle inside the box $B(r)$ as well as the time $T_r$ spent by the second particle in the same box $B(r)$. Since the time delay is a relative notion, one defines $\tau_r$ as the difference between $T_r$ and $T_r^0$. In order to have a quantity independent of the size of the boxes one finally considers the limit $\lim_{r \to \infty}\tau_r$, and says that this quantity, if it exists, is the time delay between the two scattering processes.

The above setup is obviously sensible and defines a rather comprehensible notion. However, even if these manipulations are convincing for the paradigm model, how can we generalize this procedure for a more complicated system ? Is it even possible to realize such a measure for an abstract scattering process and what is the underlying reference system ? In the same line, how can we define the notion of localisation in a box when there is no clear underlying space and no notion of boxes ?

Now, coming back to the paradigm example and assuming that the above quantity exists, one might also wonder if this quantity can be related to another measurement~? The answer is yes, and the corresponding notion is the Eisenbud-Wigner time delay \cite{Eis48,Wig55}.
In fact, the identity between the two notions of time delay
was proved in different settings by various authors (see
\cite{AC87,AJ07,AS06,BGG83,dCN02,GT07,GS80,JSM72,Jen81,Mar76,Mar81,MSA92,Rob94,RW89,
Tie06,Tie09,Wan88} and references therein), but a general and abstract statement has never been proposed.

Quite recently,  R. Tiedra de Aldecoa and the author of the present essay introduced an abstract version for the two notions of time delay and showed that the two concepts are equal \cite{BertaII}. The proof mainly relies on a general formula relating localisation operators to time
operators \cite{RT10}. Using this formula, these authors proved that the existence and the identity of the two time delays is in fact a common feature of quantum scattering
theory. Note that on the way they took into account a symmetrization procedure
\cite{AJ07,BO79,GT07,Mar75,Mar81,Smi60,Tie06,Tie08,Tie09} which broadly extends the
applicability of the theory.

The aim of the present paper is to explain how these two notions of time delay can be constructed for an abstract quantum scattering system. In particular, we introduce the necessary and natural conditions for such a construction. All assumptions and statements are precisely formulated but for the simplicity of presentation we refer to the two companion papers \cite{RT10,BertaII} for the proofs. In fact, this paper is an expansion of the presentation done by its author at the conference \emph{Spectral and scattering theory and related topics} organised in February 2011 in Kyoto in honor of Professor H. Isozaki's 60${}^{\rm th}$-Birthday.

\section{Asymptotic evolution}\label{asym}
\setcounter{equation}{0}

In this section we introduce the asymptotic system and the necessary assumptions on it.

Let $\H$ be a Hilbert space with scalar product and norm denoted respectively by $\langle \cdot, \cdot \rangle_\H$ and $\|\cdot\|_\H$. The evolution of a quantum scattering system is defined in terms of the unitary group generated by a self-adjoint operator $H$ in $\H$. One aim of scattering theory is to understand the limits as $t \to \pm \infty$ of the evolving state $\psi(t):=\e^{-itH}\psi$ for suitable $\psi\in \H$. Obviously, not all states $\psi \in \H$ can be studied and in fact only elements in the absolutely continuous subspace $\H_{\rm ac}(H)$ of $\H$ with respect to $H$ are concerned with usual scattering theory.

For investigating the long time asymptotics of $\psi(t)$ one usually looks for another Hilbert space $\H_0$ (which can also be $\H$ itself) and for a second self-adjoint operator $H_0$ in $\H_0$ such that the element $\psi(t)$ approaches for $t \to \pm \infty$ and in a suitable sense the elements $\e^{-itH_0}\psi_\pm$ for some $\psi_\pm \in \H_0$. Since in general these states do not live in the same Hilbert space, the construction requires the introduction of an operator $J:\H_0 \to \H$ usually called \emph{identification operator}. For simplicity, we shall consider $J\in \B(\H_0,\H)$ but let us mention that this boundedness condition can be relaxed if necessary.

More precisely, given the self-adjoint operator $H$ in the Hilbert space $\H$, one looks for a triple $(\H_0,H_0,J)$ such that the following strong limits exist
\begin{equation}\label{waveop}
W_\pm(H,H_0,J):=\slim_{t\to\pm\infty}\e^{itH}J\e^{-itH_0}P_{\rm ac}(H_0) \ .
\end{equation}
Assuming that the operator $H_0$ is simpler than $H$, the study of the wave operators $W_\pm(H,H_0,J)$ leads then to valuable information on the spectral decomposition of $H$. This setting is also at the root for further investigations on the evolution group generated by $H$ and in particular for our study of the time delay.

Now, let us call \emph{suitable} a triple $(\H_0,H_0,J)$ which leads to the existence of non-trivial operators $W_\pm(H,H_0,J)$ (a precise condition is stated in Assumption \ref{A3}) . Note that we are not aware of any general criterion which would insure the existence of a suitable triple. Furthermore, if any such suitable triple exists, its uniqueness can certainly not be proved. However, in the set (possibly empty) of suitable triples, the additional conditions on $H_0$ that we shall introduce in the sequel might select an optimal choice between suitable triples.

Let us recall from the Introduction that the time delay is defined in terms of expectations of evolving states on a family of \emph{position-type operators}. In an abstract setting, the existence of such a family is not guaranteed by any means and thus these operators have to be introduced by hands. So, let us assume that there exists a finite family of mutually commuting self-adjoint operators $\Phi\equiv(\Phi_1,\ldots,\Phi_d)$ in
$\H_0$ which have to satisfy two appropriate assumptions with respect to $H_0$. The first one, and by far the most important one, is a certain type of commutations relation. By looking at the examples presented in \cite[Sec.~7]{BertaII}, one can get the feeling that this assumption is related to a certain homogeneity property of an underlying configuration space. However, one has clearly not introduced such a concept up to now, and this interpretation is not based on any strong ground. The second assumption concerns the regularity of $H_0$ with respect to the operators $\Phi_1$ to $\Phi_d$. While the first assumption is easily stated, the second one necessitates some preparations.

For any $x \in \R^d$ let us set
\begin{equation*}
H_0(x):=\e^{-ix\cdot\Phi}H_0\e^{ix\cdot\Phi}
\end{equation*}
for the self-adjoint operator with domain $\e^{-ix\cdot\Phi}\dom(H_0)$.

\begin{Assumption}\label{A1}
The operators $H_0(x)$, $x \in \R^d$, mutually commute.
\end{Assumption}

Clearly, this assumption is equivalent to the commutativity of each $H_0(x)$ with $H_0$. Now, in order to express the regularity of $H_0$ with respect to $\Phi_j$, we recall from \cite[Def.~6.2.2]{ABG} that a self-adjoint operator $T$ with domain $\dom(T)\subset\H_0$ and spectrum $\sigma(T)$ is said to be of class $C^1(\Phi)$ if there exists $\omega\in\C\setminus\sigma(T)$ such that the map
\begin{equation*}
\R^d\ni x\mapsto\e^{-ix\cdot\Phi}(T-\omega)^{-1}\e^{ix\cdot\Phi}\in\B(\H_0)
\end{equation*}
is strongly of class $C^1$ in $\H_0$. In such a case and for each
$j\in\{1,\ldots,d\}$, the set $\dom(T)\cap\dom(\Phi_j)$ is a core for $T$ and the quadratic form
$
\dom(T)\cap\dom(\Phi_j)\ni\v\mapsto
\langle T\v,\Phi_j\v\rangle_{\H_0}-\langle\Phi_j\v,T\v\rangle_{\H_0}
$
is continuous in the topology of $\dom(T)$. This form extends then uniquely to a continuous quadratic form $[T,\Phi_j]$ on $\dom(T)$, which can be identified with a
continuous operator from $\dom(T)$ to its dual $\dom(T)^*$. Finally, the following equality holds:
$$
\big[\Phi_j,(T-\omega)^{-1}\big]=(T-\omega)^{-1}[T,\Phi_j](T-\omega)^{-1}.
$$
In the sequel, we shall say that $i[T,\Phi_j]$ is essentially self-adjoint on $\dom(T)$ if $[T,\Phi_j]\dom(T)\subset\H_0$ and if $i[T,\Phi_j]$ is essentially self-adjoint on $\dom(T)$ in the usual sense.

\begin{Assumption}\label{A2}
The operator $H_0$ is of class $C^1(\Phi)$, and for each $j\in\{1,\ldots,d\}$,
$i[H_0,\Phi_j]$ is essentially self-adjoint on $\dom(H_0)$, with its self-adjoint
extension denoted by $\partial_jH_0$. The operator $\partial_jH_0$ is of class
$C^1(\Phi)$, and for each $k\in\{1,\ldots,d\}$, $i[\partial_jH_0,\Phi_k]$ is
essentially self-adjoint on $\dom(\partial_jH_0)$, with its self-adjoint extension
denoted by $\partial_{jk}H_0$. The operator $\partial_{jk}H_0$ is of class
$C^1(\Phi)$, and for each $\ell\in\{1,\ldots,d\}$, $i[\partial_{jk}H_0,\Phi_\ell]$ is
essentially self-adjoint on $\dom(\partial_{jk}H_0)$, with its self-adjoint extension
denoted by $\partial_{jk\ell}H_0$.
\end{Assumption}

\begin{Remark}
Readers familiar with Mourre theory would have guessed that this assumption is closely related to a $C^3(\Phi)$-type regularity condition. However, the unusual requirement on self-adjointness is due to our use of a functional calculus associated with these successive commutators.
\end{Remark}

As shown in \cite[Sec.~2]{RT10}, this assumption implies the invariance of
$\dom(H_0)$ under the action of the unitary group $\{\e^{-ix\cdot\Phi}\}_{x\in\R^d}$.
As a consequence, each operator $H_0(x)$ has the same domain equal to $\dom(H_0)$.
Similarly, the domains $\dom(\partial_jH_0)$ and $\dom(\partial_{jk}H_0)$ are left
invariant by the action of the unitary group $\{\e^{-ix\cdot \Phi}\}_{x\in\R^d}$, and
the operators $(\partial_jH_0)(x):=\e^{-ix\cdot\Phi}(\partial_jH_0)\e^{ix\cdot\Phi}$
and $(\partial_{jk}H_0)(x):=\e^{-ix\cdot\Phi}(\partial_{jk}H_0)\e^{ix\cdot\Phi}$ are self-adjoint operators with domains $\dom(\partial_jH_0)$ and $\dom(\partial_{jk}H_0)$
respectively.
It has also been shown in \cite[Lemma~2.4]{RT10} that Assumptions \ref{A1} and \ref{A2} imply that the operators $H_0(x),(\partial_jH_0)(y)$ and $(\partial_{k\ell}H_0)(z)$
mutually commute for each $j,k,\ell\in\{1,\ldots,d\}$ and each $x,y,z\in\R^d$. For
simplicity, we set
$$
H_0':=(\partial_1H_0,\ldots,\partial_dH_0)
$$
and define for each measurable function $g:\R^d\to\C$ the operator $g(H_0')$ by using
the $d$-variables functional calculus. Similarly, we consider the family of operators $\{\partial_{jk}H_0\}$ as the components of a $d$-dimensional matrix which we denote by $H_0''$.

\begin{Remark}
By choosing for $\Phi$ the single operator $1$, or any operator commuting with $H_0$, both conditions above are satisfied by any suitable triple $(\H_0,H_0,J)$. However, as we shall see in the next section, these choices would lead to trivial statements with no information in them. In fact a criterion for an optimal choice for both $(\H_0,H_0,J)$ and $\Phi$ will be explained in Remark \ref{optimal}.
\end{Remark}

We are already in a suitable position for the definition of the sojourn time for the evolution group generated by $H_0$. However, we would like first to look at various consequences on $H_0$ of the Assumptions \ref{A1} and \ref{A2}.

\section{Properties of $\boldsymbol{H_0}$}\label{SurH_0}
\setcounter{equation}{0}

In this section we assume tacitly that Assumptions \ref{A1} and \ref{A2} hold and exhibit some consequences on the operator $H_0$. Our first task is to define values in the spectrum of $H_0$ which have a troublesome behaviour for scattering theory. Obviously, this set can only be defined with the objects yet at hand.

For that purpose, let $E^{H_0}(\;\!\cdot\;\!)$ denote the spectral measure of $H_0$. For shortness, we also use the notation $E^{H_0}(\lambda;\delta)$ for
$E^{H_0}\big((\lambda-\delta,\lambda+\delta)\big)$.
We now introduce the set of critical values of $H_0$ and state its main properties, see also \cite[Lemma~2.6]{RT10} for more properties and details.

\begin{Definition}\label{surkappa}
A number $\lambda\in\sigma(H_0)$ is called a critical value of $H_0$ if
\begin{equation*}
\lim_{\varepsilon\searrow0}\big\|\big((H_0')^2+\varepsilon\big)^{-1}
E^{H_0}(\lambda;\delta)\big\|_{\H_0}=+\infty
\end{equation*}
for each $\delta>0$. We denote by $\kappa(H_0)$ the set of critical values of $H_0$.
\end{Definition}

\begin{Lemma}\label{Heigen}
Let $H_0$ satisfy Assumptions \ref{A1} and \ref{A2}. Then the set
$\kappa(H_0)$ is closed and contains the set of eigenvalues of $H_0$.
\end{Lemma}

Now, the spectral properties of $H_0$ which are exhibited in the next proposition are consequences of the existence of an explicit conjugate operator for $H_0$. Indeed, it has been shown in \cite[Sec.~3]{RT10} that for $j \in \{1,\dots,d\}$ the expression
$\Pi_j:=\langle H_0\rangle^{-2} (\partial_j H_0) \langle H_0\rangle^{-2}$ defines a bounded self-adjoint operator and that the operators $\Pi_j$ and $\Pi_k$ commute for arbitrary $j,k$.
Note that we use the notation
$\langle x\rangle:=(1+x^2)^{1/2}$ for any $x\in\R^n$.
Based on this, it is proved in the same reference that the operator
\begin{equation*}
A:=\frac{1}{2}\big(\Pi\cdot \Phi + \Phi\cdot \Pi\big)
\end{equation*}
is essentially self-adjoint on the domain $\dom(\Phi^2)$.
Then, since the formal equality
$$
[iH_0,A]=\langle H_0\rangle^{-2} (H_0')^2 \langle H_0\rangle^{-2}
$$
holds, the commutator $[iH_0,A]$ is non-negative and this construction opens the way to the study of the operator $H_0$ with the so-called Mourre theory. Such an analysis has been performed in \cite[Sec.~3]{RT10} from which we recall the main spectral result:

\begin{Proposition}\label{not_bad}
Let $H_0$ satisfy Assumptions \ref{A1} and \ref{A2}. Then,
\begin{enumerate}
\item[(a)] the spectrum of $H_0$ in $\sigma(H_0)\setminus\kappa(H_0)$ is purely
absolutely continuous,
\item[(b)] each operator $B\in\B\big(\dom(\langle\Phi\rangle^{-s}),\H_0\big)$, with
$s>1/2$, is locally $H_0$-smooth on $\R\setminus\kappa(H_0)$.
\end{enumerate}
\end{Proposition}

\begin{Remark}
It is worth noting that modulo the regularization $\langle H_0\rangle^{-2}$, the usual conjugate operator for the Laplace operator $\Delta$ in $\ltwo(\R^d)$ is constructed similarly. Indeed, if we choose for $\Phi$ the family of position operators $X=(X_1,\dots,X_d)$, then Assumptions \ref{A1} and \ref{A2} are clearly satisfied and the generator of dilation is obtained by this procedure.
\end{Remark}

\begin{Remark}\label{optimal}
One would like to stress that the definition of the set $\kappa(H_0)$ clearly depends on the choice of the family of operators $\Phi = \{\Phi_1,\dots,\Phi_d\}$. For example if $\Phi = \{1\} $, then $H_0' = 0$ and $\kappa(H_0)=\sigma(H_0)$, and it follows that Proposition \ref{not_bad} does not contain any information. Thus the choice for both a suitable triple $(\H_0,H_0,J)$ and the family of operators $\Phi$ should be dictated by the size of the corresponding set $\kappa(H_0)$: the smaller the better.
\end{Remark}

\section{Sojourn times and symmetrized time delay}\label{SecSymDelay}
\setcounter{equation}{0}

In this section we introduce the notions of sojourn times for the two evolution groups and define the symmetrized time delay. We also state the main result on the existence of the symmetrized time delay under suitable assumptions on the scattering system.
But first of all, let us state the precise assumption on a triple $(\H_0,H_0,J)$ for being suitable. More precisely, this assumption concerns the existence, the isometry and the completeness of the generalised wave operators.

\begin{Assumption}\label{A3}
The generalised wave operators $W_\pm(H,H_0,J)$ defined in \eqref{waveop} exist and are partial isometries with final subspaces $\H_{\rm ac}(H)$.
\end{Assumption}

The initial subspaces of the wave operators are denoted by $\H_0^\pm \subset\H_{\rm ac}(H_0)$.
In fact, it follows from a standard argument that the operator $H_0$  is reduced by the decompositions $\H_0^\pm \oplus \big(\H_0^\pm\big)^\bot$ of $\H_0$, {\it cf.}~\cite[Prop.~6.19]{BauWol}. Furthermore, the main consequence of Assumption \ref{A3} is that the scattering operator
\begin{equation*}
S:= W_+^*W_-:\H_0^-\to\H_0^+
\end{equation*}
is a well-defined unitary operator commuting with $H_0$. Note that if $S$ is considered from $\H_0$ into itself, then this operator is only a partial isometry, with initial subset $\H_0^-$ and final subset $\H_0^+$.

We now define the sojourn times for the quantum scattering system,
starting with the sojourn time for the free evolution $\e^{-itH_0}$. For that purpose, let us first define for any $s\geq 0$
$$
\D_s:=\big\{\varphi\in\dom\big(\langle \Phi \rangle^s\big) \mid\varphi=\eta(H_0)\varphi
\hbox{ for some }\eta\in C^\infty_{\rm c}\big(\R\setminus\kappa(H_0)\big)\big\}\ .
$$
The set $\D_s$ is included in the subspace $\H_{\rm ac}(H_0)$ of absolute
continuity of $H_0$, due to Proposition \ref{not_bad}.(a), and
$\D_{s_1}\subset\D_{s_2}$ if $s_1\ge s_2$.
We also refer the reader to \cite[Sec.~6]{RT10} for an account on density properties of the sets $\D_s$.
Then, let $f$ be a non-negative even element of the Schwartz space $\S(\R^d)$ equal to $1$ on a neighbourhood $\Sigma$ of the origin $0\in\R^d$. Here even means that $f(-x)=f(x)$ for any $x \in \R^d$. For $r>0$ and $\varphi\in\D_0$, we set
$$
T_r^0(\varphi)
:=\int_\R\d t\,\big\langle\e^{-itH_0}\varphi,f(\Phi/r)\e^{-itH_0}\varphi\big\rangle_{\H_0},
$$
where the integral has to be understood as an improper Riemann integral. The operator
$f(\Phi/r)$ is approximately the projection onto the subspace $E^\Phi(r\Sigma)\H_0$
of $\H_0$, with $r\Sigma:=\{x\in\R^d\mid x/r\in\Sigma\}$. Therefore, if
$\|\varphi\|_{\H_0}=1$, then $T_r^0(\varphi)$ can be approximately interpreted as the time
spent by the evolving state $\e^{-itH_0}\varphi$ inside $E^\Phi(r\Sigma)\H_0$.
Furthermore, the expression $T_r^0(\varphi)$ is finite for each $\varphi\in\D_0$,
since we know from Proposition \ref{not_bad}.(b) that each operator
$B\in\B\big(\dom(\langle\Phi\rangle^{-s}),\H_0\big)$, with $s>\12$, is locally
$H_0$-smooth on $\R\setminus\kappa(H_0)$.

When trying to define the sojourn time for the full evolution $\e^{-itH}$, one faces the
problem that the localisation operator $f(\Phi/r)$ acts in $\H_0$ while the operator
$\e^{-itH}$ acts in $\H$. The obvious modification would be to consider the operator
$Jf(\Phi/r)J^*\in\B(\H)$, but the resulting framework could be not general enough. Sticking to the basic idea that the freely evolving
state $\e^{-itH_0}\varphi$ should approximate, as $t\to\pm\infty$, the corresponding
evolving state $\e^{-itH}W_\pm\varphi$, one looks for an operator
$L(t):\H\to\H_0$, $t\in\R$, such that
\begin{equation}\label{defconv}
\lim_{t\to\pm\infty}\left\|L(t)\e^{-itH}W_\pm\varphi-\e^{-itH_0}\varphi\right\|_{\H_0}=0.
\end{equation}
Since we consider vectors $\varphi\in\D_0$, the operator $L(t)$ can be unbounded as
long as $L(t)E^{H}(I)$ is bounded for any bounded subset $I\subset\R$. With such a
family of operators $L(t)$, it is natural to define a first contribution for the sojourn time of the full
evolution $\e^{-itH}$ by the expression
\begin{equation*}
T_{r,1}(\varphi):=\int_\R\d t\,
\big\langle \e^{-itH}W_-\varphi,L(t)^*f(\Phi/r)L(t)\e^{-itH}W_-\varphi\big\rangle_\H.
\end{equation*}
However, another contribution naturally appears in this context. Indeed, for fixed $t$, the localisation operator $L(t)^*f(\Phi/r)L(t)$ strongly converges to $L(t)^*L(t)$ as $r \to \infty$, but this operator might be rather different from the operator $1$. As a consequence, a part of the Hilbert space might be not considered with the definition of $T_{r,1}(\varphi)$. Thus, a second contribution for the sojourn time is
\begin{equation*}
T_2(\varphi):=\int_\R\d t\,\big\langle\e^{-itH}W_-\varphi,\big(1-L(t)^*L(t)\big)
\e^{-itH}W_-\varphi\big\rangle_\H.
\end{equation*}

The finiteness of $T_{r,1}(\varphi)$ and $T_2(\varphi)$ is proved under an additional
assumption in Theorem \ref{exist} below. The term $T_{r,1}(\varphi)$ can be
approximatively interpreted as the time spent by the scattering state
$\e^{-itH}W_-\varphi$ inside $L(t)^*f(\Phi/r)L(t)\H$.
The term $T_2(\varphi)$ can be seen as the time spent by the scattering state
$\e^{-itH}W_-\varphi$ inside the time-dependent subset $\big(1-L(t)^*L(t)\big)\H$ of
$\H$. If $L(t)$ is considered as a time-dependent quasi-inverse for the
identification operator $J$ (see \cite[Sec.~2.3.2]{Yaf92} for the related
time-independent notion of quasi-inverse), then the subset $\big(1-L(t)^*L(t)\big)\H$
can be seen as an approximate complement of $J\H_0$ in $\H$ at time $t$. When
$\H_0=\H$, one usually sets $L(t)=J^*=1$, and the term $T_2(\varphi)$ vanishes. Within
this general framework, the total sojourn time for the full
evolution $\e^{-itH}$ is given by
$$
T_r(\varphi):= T_{r,1}(\varphi) + T_2(\varphi)\ .
$$

Since both sojourn times have now been defined, the definition of the time delay should be at hand. However, let us first consider the following dilemma. For a given state $L(t)\e^{-itH}\psi$ with $\psi \in \H_{\rm ac}(H)$, which one is the correct free evolution state : is it $\e^{-itH_0}\varphi_-$ with $W_-\varphi_- = \psi$ which is a good approximation for $t \to -\infty$, or is it $\e^{-itH_0}\varphi_+$ with $W_+\varphi_+ = \psi$ which is also a good approximation but for $t \to +\infty$ ?
Obviously, both states have to be taken into account, and therefore  we say that
\begin{equation*}
\tau_r(\varphi):=T_r(\varphi)-\12\big\{T_r^0(\varphi)+T_r^0(S\varphi)\big\},
\end{equation*}
is the symmetrized time delay of the scattering system with incoming state $\varphi$. This symmetrized
version of the usual time delay
\begin{equation*}
\tau_r^{\rm in}(\varphi):=T_r(\varphi)-T_r^0(\varphi)
\end{equation*}
is known to be the only time delay having a well-defined limit as $r\to\infty$ for
complicated scattering systems (see for example
\cite{AJ07,BO79,GT07,IRT11,Mar75,Mar81,BertaII,SM92,Smi60,Tie06}).

The last assumption is a condition on the speed of convergence of the state $L(t)\e^{-itH}W_\pm \varphi_\pm$ to the corresponding states $\e^{-itH_0}\varphi_\pm$ as $t \to \pm \infty$. Up to now, only the convergence to $0$ of the norm of the difference of these states had been used, {\it cf.}~\eqref{defconv}.

\begin{Assumption}\label{A4}
For each $\varphi_\pm\in\H_0^\pm\cap\D_0$ one has
\begin{equation}\label{H-+}
\big\|\big(L(t)W_--1\big)\e^{-itH_0}\varphi_-\big\|_{\H_0}\in\lone(\R_-,\d t)
\ \ \hbox{and}\ \ \big\|(L(t)W_+-1)\e^{-itH_0}\varphi_+\big\|_{\H_0}\in\lone(\R_+,\d t).
\end{equation}
\end{Assumption}

Next Theorem shows the existence of the symmetrized time delay. The apparently large number of assumptions reflects nothing more
but the need of describing the very general scattering system; one needs hypotheses on the relation between $H_0$ and $\Phi$, a compatibility assumption between $H_0$ and $H$, conditions on the localisation function $f$ and conditions on the
state $\varphi$ on which the calculations are performed.

\begin{Theorem}\label{exist}
Let $H$, $H_0$ $J$ and $\Phi$ satisfy Assumptions \ref{A1} to $\ref{A4}$, and let $f$ be a non-negative even element of $\S(\R^d)$ equal to $1$ on a neighbourhood of the origin $0\in\R^d$.
Then, for each $\varphi\in\H_0^-\cap\D_2$ satisfying
$S\varphi\in\D_2$, the sojourn time $T_r(\varphi)$ is finite for each $r>0$ and the limit $\lim_{r\to\infty}\tau_r(\varphi)$ exists.
\end{Theorem}

\begin{Remark}
All the assumptions in the above statement are rather explicit except the one on $S\varphi \in \D_2$. Indeed, such a property is related to the mapping properties of the scattering operator and this assumption is not directly connected to the other conditions. Let us simply mention that one usually proves such a property by studying higher order resolvent estimates.
\end{Remark}

In the next section, we show that the time delay $\lim_{r\to\infty}\tau_r(\varphi)$ can be related to another quantity defined only in terms of the scattering operator and a so-called time operator.

\section{Time operator and Eisenbud-Wigner time delay}\label{Time_operator}
\setcounter{equation}{0}

We now define a time operator for the operator $H_0$ and recall some of its properties from \cite{RT10}. For that purpose, one needs to construct a new function $R_f$ from the localisation function $f$ introduced above. This function was already studied and used, in one form or another, in \cite{GT07,RT10,BertaII,Tie08,Tie09}.
Thus, let us define $R_f \in C^\infty\big(\R^d\setminus\{0\}\big)$  by
$$
R_f(x):=\int_0^\infty\frac{\d\mu}\mu\big(f(\mu x)-\chi_{[0,1]}(\mu)\big)\ .
$$
The following properties of $R_f$ are proved in
\cite[Sec. 2]{Tie09}:
$R_f'(x)=\int_0^\infty\d\mu\;\!f'(\mu x)$,
$x\cdot R_f'(x)=-1$ and
$t^{|\alpha|}(\partial^\alpha R_f)(tx)=(\partial^\alpha R_f)(x)$, where
$\alpha\in\N^d$ is a multi-index and $t>0$. Furthermore, if $f$ is radial, then $R_f'(x)=-x^{-2}x$.

Now, the next statement follows from \cite[Prop.~5.2]{RT10} and
\cite[Rem.~5.4]{RT10}.

\begin{Proposition}\label{lemma_T_f}
Let $H_0$ and $\Phi$ satisfy Assumptions \ref{A1} and \ref{A2}, and let $f$ be the localisation function introduced above. Then the map
$$
t_f:\D_1\to\C,\quad\v\mapsto
t_f(\v):=-\12\sum_{j=1}^d\big\{\big\langle\Phi_j\v,(\partial_jR_f)(H_0')\v\big\rangle_{\H_0}
+\big\langle\big(\partial_jR_f \big)(H_0')\v,\Phi_j\v\big\rangle_{\H_0}\big\},
$$
is well-defined. Moreover, the linear operator $T_f:\D_1\to\H_0$ defined by
\begin{equation}\label{nemenveutpas}
\textstyle
T_f\v:=-\12\big(\Phi\cdot R_f'(H_0')+R_f'\big(\frac{H_0'}{|H_0'|})\cdot\Phi\;\!|H_0'|^{-1}
+iR_f'\big(\frac{H_0'}{|H_0'|}\big)\cdot\big(H_0''^{\sf T}H_0'\big)|H_0'|^{-3}\big)\v
\end{equation}
satisfies $t_f(\v)=\langle\v,T_f\v\rangle$ for each $\v\in\D_1$. In particular,
$T_f$ is a symmetric operator if $\D_1$ is dense in $\H_0$.
\end{Proposition}

Clearly, Formula \eqref{nemenveutpas} is rather complicated and one could be tempted to replace it by the simpler formula
$-\12\big(\Phi\cdot R_f'(H_0')+R_f'(H_0')\cdot\Phi\big)\varphi$. Unfortunately, a precise
meaning of this expression is not available in general, and its full derivation can only be justified in concrete examples.

Before stating the main result of this section, let us recall some properties of the operator $T_f$, and refer to \cite[Sec.~6]{RT10} for details. In the form sense on $\D_1$ the operators $H_0$ and $T_f$ satisfy the canonical commutation relation
$$
[T_f,H_0]=i.
$$
Therefore, since the group $\{\e^{-itH_0}\}_{t\in\R}$ leaves $\D_1$ invariant, the
following equalities hold in the form sense on $\D_1$:
\begin{equation*}
\textstyle\big\langle\psi,T_f\e^{-itH_0}\v\big\rangle_{\H_0}
=\big\langle\psi,\e^{-itH_0}\big(T_f+t\big)\v\big\rangle_{\H_0}\ ,
\end{equation*}
and the operator $T_f$ satisfies on $\D_1$ the
so-called infinitesimal Weyl relation in the weak sense \cite[Sec.~3]{JM80}. Note
that we have not supposed that $\D_1$ is dense. However, if $\D_1$ is dense in
$\H_0$, then the infinitesimal Weyl relation in the strong sense holds:
\begin{equation}\label{strongWeyl}
\textstyle
T_f\e^{-itH_0}\v=\e^{-itH_0}\big(T_f+t\big)\v,\qquad\v\in\D_1.
\end{equation}
This relation, also known as $T_f\,$-weak Weyl relation \cite[Def.~1.1]{Miy01},
has deep implications on the spectral nature of $H_0$ and on the form of $T_f$ in
the spectral representation of $H_0$. Formally, it suggests that
$T_f=i\frac\d{\d H_0}$, and thus $-iT_f$ can be seen as the operator of
differentiation with respect to the Hamiltonian $H_0$. Moreover, being a weak
version of the usual Weyl relation, Relation \eqref{strongWeyl} also suggests
that the spectrum of $H_0$ may not differ too much from a purely absolutely
continuous spectrum. Since these properties have been thoroughly discussed in \cite[Sec.~6]{RT10}, we refer the interested reader to that reference.

Next theorem is the main result of \cite{BertaII}, comments on it are provided after its statement.

\begin{Theorem}\label{sym_case}
Let $H$, $H_0$ $J$ and $\Phi$ satisfy Assumptions \ref{A1} to $\ref{A4}$, and let $f$ be a non-negative even element of $\S(\R^d)$ equal to $1$ on a neighbourhood of the origin $0\in\R^d$.
Then, for each $\varphi\in\H_0^-\cap\D_2$ satisfying
$S\varphi\in\D_2$ one has
\begin{equation}\label{Eisenbud_sym}
\lim_{r\to\infty}\tau_r(\varphi)
=-\big\langle\varphi,S^*\big[T_f,S\big]\varphi\big\rangle_{\H_0},
\end{equation}
with $T_f$ defined by \eqref{nemenveutpas}.
\end{Theorem}

The above statement expresses the identity of
the symmetrized time delay (defined in terms of sojourn times) and the Eisenbud-Wigner
time delay for general scattering systems. The l.h.s. of
\eqref{Eisenbud_sym} is equal to the symmetrized time delay of the scattering system with incoming state $\varphi$, in the dilated regions associated with the localisation operators $f(\Phi/r)$. The r.h.s. of \eqref{Eisenbud_sym} is the
expectation value in $\varphi$ of the generalised Eisenbud-Wigner time delay operator $-S^*[T_f,S]$.
It clearly shows that once suitable and natural conditions are assumed, then the notion of time delay exists whatever the scattering system is.

Let us finally mention that when $T_f$ acts in the spectral representation of $H_0$ as the
differential operator $i\frac\d{\d H_0}$, which occurs in most of the situations of
interest (see for example \cite[Sec.~7]{RT10}), one recovers the usual Eisenbud-Wigner
Formula:
$$
\lim_{r\to\infty}\tau_r(\varphi)
\textstyle=-\big\langle\varphi,iS^*\frac{\d S}{\d H_0}\;\!\varphi\big\rangle_{\H_0}.
$$


\end{document}